\documentclass[twocolumn,preprintnumbers,superscriptaddress,aps,prd,floatfix,nofootinbib]{revtex4-1}

\usepackage{enumerate}
\usepackage{amsmath}
\usepackage{graphicx}
\usepackage{placeins}
\usepackage{xspace,slashed}
\usepackage{hyperref}
\usepackage{braket}
\hypersetup{colorlinks=true, citecolor=blue, urlcolor=blue, linkcolor=blue}
\usepackage[normalem]{ulem}
\usepackage{booktabs}

\usepackage[capitalize]{cleveref}
\usepackage{siunitx}

\renewcommand{\vec}[1]{{\bf{#1}}}

\newcommand{\program}[1]{\texttt{#1}\xspace}
\newcommand{\mg}{\program{MadGraph5\_aMC@NLO}}
\newcommand{\smeftsim}{\program{SMEFTsim}}
\newcommand{\eos}{\program{EOS}}
\newcommand{\flavio}{\program{flavio}}
\newcommand{\smelli}{\program{smelli}}
\newcommand{\wilson}{\program{wilson}}
\newcommand{\dsixtools}{\program{DSixTools}}
\newcommand{\smefit}{{\sc\small SMEFiT}}
\newcommand{\fitmaker}{\program{Fitmaker}}
\newcommand{\version}[1]{\texttt{v#1}\xspace}

\begin{document}

\providecommand{\abs}[1]{\lvert#1\rvert}

\preprint{IPPP/24/72}
\preprint{SI-HEP-2024-24}
\preprint{P3H-24-086}

\title{Collider-Flavour Complementarity from the bottom to the top}
\begin{abstract}
Motivated by recently observed anomalies in the flavour sector, we analyse the potential of measurements of top quarks at the Large Hadron Collider (LHC) to provide complementary constraints on interactions that shape low-energy precision investigations in the $B$ sector. The measurement of top quark properties, such as the top width and the abundant top pair production channels, are already reaching the percent level at this relatively early stage of the LHC phenomenology program.
A focused analysis of four-fermion interactions, employing effective field theory without flavour structure assumptions and incorporating renormalization group evolution effects, bridges $B$ meson scale phenomena with key top quark measurements. We demonstrate that the LHC is increasingly competitive with, and complementary to, flavour physics constraints.
Our results, which include a first comprehensive analysis of non-leptonic $B$ decays in this context, suggest that the LHC’s top physics program could serve as a valuable, complementary tool in the search for physics beyond the Standard Model within the flavour sector.
\end{abstract}
\author{Oliver Atkinson} 
\affiliation{School of Physics and Astronomy, University of Glasgow, Glasgow G12 8QQ, UK\\[0.10cm]}
\author{Christoph Englert} \email{christoph.englert@glasgow.ac.uk}
\affiliation{School of Physics and Astronomy, University of Glasgow, Glasgow G12 8QQ, UK\\[0.10cm]}
\author{Matthew Kirk} \email{matthew.j.kirk@durham.ac.uk}
\affiliation{Institute for Particle Physics Phenomenology, Durham University, Durham, DH1 3LE, United Kingdom\\[0.10cm]}
\author{Gilberto Tetlalmatzi-Xolocotzi} \email{gtx@physik.uni-siegen.de}
\affiliation{Physik Department, Universit\"{a}t Siegen, Walter-Flex-Str. 3, 57068 Siegen, Germany\\[0.10cm]
}
\affiliation{Universit\`e Paris-Saclay, CNRS/IN2P3, IJCLab, 91405 Orsay, France\\[0.1cm]}

\flushbottom
\allowdisplaybreaks
\maketitle

\section{Introduction}
\label{sec:intro}
Some ten years after the Higgs boson discovery, beyond the Standard Model (BSM) physics remains elusive in the high-energy collisions observed at the Large Hadron Collider (LHC). While measurements since 2013 have, on the one hand, consolidated the Standard Model (SM) as a surprisingly accurate description of the weak gauge-Higgs sector, a range of flavour physics measurements call the SM into question. These anomalies, chiefly parametrized using the language of effective field theory (EFT)~\cite{Weinberg:1978kz} are based on measurements of the bottom sector and have triggered a range of theoretical investigations into the potential UV origins of the observed deviations.

EFT is an established tool to investigate the (dis)agreement of theoretical predictions with experimental findings in flavour physics. Its application to the multi-scale processes in the LHC environment, however, is relatively recent. Motivated by the, so far, unsuccessful searches for concrete and motivated BSM extensions, EFT methods are increasingly becoming a new standard for communicating results. Owing to the electroweak precision constraints from the LEP era and the broadly observed consistency of Higgs interactions with the SM-predicted patterns, a typical focus of these investigations is Standard Model effective field theory (SMEFT)~\cite{Grzadkowski:2010es}. SMEFT is based on the gauge symmetry and field content of the SM and describes the leading deformations of the SM interactions. The Lagrangian in this approximation is schematically given as
\begin{equation}
\mathcal{L}_{\text{BSM}} = \mathcal{L}_{\text{SM}} + \sum_i \frac{{C}_i}{\Lambda^2} {Q}_i \,,
\end{equation}
and is characterized by a cut-off suppression $\sim \Lambda^{-2}$. The scale and scheme-dependent Wilson coefficients ${{C}}_i$ can be related to concrete UV scenarios through matching calculations, which have received substantial attention recently~\cite{DasBakshi:2018vni,Carmona:2021xtq,Fuentes-Martin:2022jrf}. Efforts to holistically constrain SMEFT at the LHC are well underway; proof-of-principle investigations in a range of phenomenological arenas have been provided in Refs.~\cite{Buckley:2015nca,Buckley:2015lku,Brivio:2019ius,Hartland:2019bjb,Banerjee:2020vtm,Ellis:2020unq,Ethier:2021bye,Aoude:2022deh,Elmer:2023wtr}. SMEFT deformations at different scales are dominantly related by renormalization group running effects which have been detailed in~Refs.~\cite{Jenkins:2013zja,Jenkins:2013wua,Alonso:2013hga,Elias-Miro:2013mua,Celis:2017hod} (see also~\cite{Greljo:2023bdy}).

In this work, we consider the recently arising puzzle in the context of the annihilation-free non-leptonic $B$ meson decays into heavy-light final states~\cite{Bordone:2020gao} (see in particular the very recent global analysis of \cite{Meiser:2024zea}) from a perspective of flavour-collider complementarity. This also offers an ideal context for drawing broader conclusions about the opportunities that come to light from comparing and combining high-precision flavour measurements with high-energy collider experiments (see in particular Refs.~\cite{Basler:2016obg,Bissmann:2019qcd,Bissmann:2019gfc,Bissmann:2020mfi,Fuentes-Martin:2020lea,Bordone:2021cca,Atkinson:2021eox,Bruggisser:2021duo,Bruggisser:2022rhb,Atkinson:2022pcn,Bhattacharya:2023beo,Grunwald:2023nli,Englert:2024nlj}).

We start by providing a brief description of the non-leptonic puzzle. A well-established technique for the calculation of non-leptonic $B$ meson decays is QCD-factorization (QCDF)~\cite{Beneke:1999br,Beneke:2000ry,Beneke:2001ev,Beneke:2003zv,Huber:2016xod}, which has been developed up to next-to-next-to leading order (NNLO) in the $\alpha_s$ expansion~\cite{Beneke:2006mk,Pilipp:2007mg,Bell:2007tv,Bell:2009nk,Beneke:2009ek,Bell:2009fm, Bell:2015koa,Bell:2020qus}. Unfortunately, most non-leptonic $B$ meson decays suffer from large uncertainties due to the presence of non-factorizable contributions. However, there are a subset of decays into heavy-light final states $B\rightarrow D^{(*)} L$ (which feature a light $L=\pi, K$ meson in addition to the $D$~\cite{Huber:2016xod}) for which it is possible to identify specific processes where non-factorizable contributions due to annihilation topologies are absent. Phenomenologically important examples are $\bar{B}^0\rightarrow D^+ K^{(*)-}$ and $\bar{B}^0\rightarrow D_s^{(*)+} \pi^{-}$. Interestingly, recent updates in the determination of different observables for these decays have revealed considerable tension between theory and experiment~\cite{Bordone:2020gao}. In light of the absence of the hadronic uncertainties that affect most of the two body non-leptonic $B$ meson decays, these deviations are particularly inviting.

There are two possible explanations. The first one is that our understanding of the dynamics of hadronic effects is less robust than expected, as our theoretical tools fail to provide a fair description of the experimental results in the cleanest category of non-leptonic $B$ meson decays. The second alternative is that these deviations may, indeed, indicate the presence of new physics (NP). This possibility singles out a well-defined subset of BSM four-quark operators, which can be matched onto SMEFT operators. We target these directions in the EFT parameter space with an analysis that employs a multi-scale approach across the naive flavour-collider divide. 

The focus of our collider investigation will be top quarks. These are immediately motivated final states, given their weak left-chiral relation with the bottom sector, since in the SMEFT the relevant degrees of freedom are left-chiral weak doublets.\footnote{Counter examples to this motivation can be constructed, and the right-handed sector can be more easily modified in reflection of that. Nonetheless, many motivated UV extensions of the SM feature the left-handed doublet character to scales above the TeV scale.}
In addition, their phenomenology enables us to illustrate how the high precision that is achieved in LHC measurements is {\emph{already now}} becoming competitive with constraints arising from flavour physics measurements at lower energy.
In contrast to many earlier analyses, we make no assumptions about the flavour structure in the SMEFT.
Connecting the different energy scales probed in top quark and $B$ physics analyses under well-defined assumptions is then critical to further or relieve the observed tension of the different data sets synergetically.

We organize this work as follows. Sec.~\ref{sec:flavour} details our flavour physics analysis: we summarize the basic formalism in QCDF, which is relevant to this project, before discussing relevant flavour observables and their implementation into our limit setting. Then, in Sec.~\ref{sec:lhc}, we discuss how to connect our low energy $B$ physics with LHC measurements and present representative top-quark measurements from the LHC. In Sec.~\ref{sec:interp}, we then move to comparing flavour and collider phenomenology results, highlighting points of complementarity between the two measurement realms. We conclude in Sec.~\ref{sec:conc}.

\section{Flavour physics constraints}
\label{sec:flavour}
\subsection{Non-leptonic B decay analysis}
\label{sec:bdecay}
\subsubsection*{Annihilation-free non-leptonic channels}
Non-leptonic processes suffer, in general, from significant theoretical uncertainties. For example, within the framework of QCDF, a first-principle determination of annihilation topologies (which are crucial to most two-body decay channels) is not feasible, see \cite{Piscopo:2023opf} for an alternative procedure on the calculation of non-leptonic $B$ meson decays. This is due to multiple sources of infrared divergences that arise particularly in transitions involving light meson pairs in the final state, such as $B\rightarrow \pi\pi$, $B\rightarrow \pi K$ and $B\rightarrow K K$ among others.
However, such sources of uncertainty are absent in some heavy-light two-body decays such as $\bar{B}^0\rightarrow D^+ K^{(*)-}$ and  $\bar{B}^0\rightarrow D_s^{(*)+} \pi^{-}$~\cite{Bordone:2020gao}.

The QCDF computations follow an effective theory approach; the set of four-quark operators that are relevant for the scope of our work are of the form $Q = (\bar{c} \Gamma b) (\bar{q} \Gamma u)$~\cite{Buras:2000if, Buras:2012gm} where there are two potential colour contractions, the Dirac structures are of the form $\gamma_\mu P_L \otimes \gamma^\mu P_L$, $\gamma_\mu P_L \otimes \gamma^\mu P_R$, $P_L \otimes P_L$, $P_L \otimes P_R$, or $\sigma_{\mu \nu} P_L \otimes \sigma^{\mu \nu} P_L$ (plus chirality flipped conjugates), and $q = \{d,s\}$, leading to a total of 40 operators, of which only the $\gamma_\mu P_L \otimes \gamma^\mu P_L$ structure is present in the SM.

Within the formalism of QCDF, the matrix elements of these operators obey the following generic decomposition~\cite{Beneke:2000ry}
\begin{multline}
    \bra{D_q^{(*)+} L^-} Q_i \ket{\bar{B}_q^0}
        =  \sum_{j} F_j^{\bar{B}_q \to D_q^{(*)}}(M_L^2) \\
        \times\int_0^1 \text{d}u\, T_{ij}(u) \Phi_L(u)+ \mathcal{O}\left(\frac{\Lambda_\text{QCD}}{m_b}\right)\,,
\end{multline}
where $T_{ij}$ are the hard scattering kernels which can be computed perturbatively. $\Phi_L$ refers to the light-cone distribution amplitude for the meson $L$, and $F_j^{\bar{B}_q \to D_q^{(*)}}$ are the relevant form factors to describe the transitions $\bar{B}_q \to D_q^{(*)}$. For the computation of the matrix elements, we use the calculations provided in~Ref.~\cite{Cai:2021mlt} after independent cross checks. The physical amplitudes can be written as
\begin{equation}
	{\mathcal{A}}(\bar{B}_{(s)}^0 \to D_{(s)}^{(*)+} L^-) = A_{D_{(s)}^{(*)+} L^-}\,\left[a_1(D_{(s)}^{(*)+} L^-)\right]\,,
\end{equation}
where the prefactor $A_{D_{(s)}^{(*)+} L^-}$ collects the relevant CKM structure, the form factor $F_j^{B \to D^{(*)}}$  for the transition $B\rightarrow D^{(*)}$ and the decay constant $f_{L}$ of the light meson $L$, while $a_1$ represents the perturbative calculation in terms of EFT Wilson coefficients.
The SM part of $a_1(D_{(s)}^{(*)+} L^-)$ has been calculated up to NNLO in the $\alpha_s$ expansion employing QCDF in~Ref.~\cite{Huber:2016xod}. In~\cite{Cai:2021mlt}, the perturbative kernels corresponding to the BSM contributions have been determined up to NLO within QCDF; we deploy these results for our analysis.

\subsubsection*{B-physics observables}
Central to our study are the constructed observables~\cite{Bjorken:1988kk}
\begin{equation}
\begin{split}
	R_{(s)L}^{(\ast)} &\equiv \frac{\Gamma(\bar{B}_{(s)}^0\to D_{(s)}^{(\ast)+}L^-)}{\text{d}\Gamma(\bar{B}_{(s)}^0\to D_{(s)}^{(\ast)+}\ell^-\bar{\nu}_{\ell})/\text{d}q^2\mid_{q^2=m_L^2}}\\\ 
	&=6\pi^2\,|V_{uq}|^2\,f_L^2\,|a_1(D_{(s)}^{(\ast)+}L^-)|^2\, X_{(s)L}^{(\ast)}\,.
\label{eq:ratios}	
\end{split}
\end{equation}
This construction ensures that the $V_{cb}$ dependence vanishes (which is helpful in light of ongoing tensions between inclusive and exclusive determinations, see the PDG review~\cite{ParticleDataGroup:2024cfk,PDG:2024VcbVub} for a summary), as well reducing the form-factor dependence, since the $X_L$ factors are ratios of the required form factors (the general definition can be found in~\cite{Neubert:1997uc} for pseudoscalar and vector mesons $D^{(*)}$). Numerically we evaluate these form-factor ratios using the software \eos~\cite{EOSAuthors:2021xpv}, which enables us to incorporate state-of-the-art results while accounting for correlations, which leads to low total uncertainties for the different quantities $X^{(*)}_L$ (following Ref.~\cite{Neubert:1997uc} we consider $X_{K^{*}}=1$):
\begin{equation}
\begin{aligned}
X_{\pi} &= 1.0012000(1) \,,  & X_{(s)\pi} &= 1.00111(8) \,, \\
X^{*}_{K} &= 0.944(5) \,, & X^{*}_{(s)\pi} &= 0.945(8) \,.
\end{aligned}
\label{eq:Xfactors}
\end{equation}
These precise results mean the ratios $R^{(*)}_{(s)L}$ show very low sensitivity to hadronic uncertainties arising from the form factors. Thus in view of the high precision of the CKM elements $V_{uq}$, the leading uncertainties in the theoretical determination of $R^{(*)}_{(s)L}$ stem from the decay constant $f_L$ and the renormalization scale uncertainty which affects $a_{1}(D^{(*)+}_{(s)}L^{-})$.
%
\begin{table}[!t]
\renewcommand{\arraystretch}{1.7}  
\begin{tabular}{ c| c }
\hline
Observable  &  Experimental value \\ 
 \hline
$\mathrm{Br}(\bar{B}^0\rightarrow D^+ K^-)$ & $(2.05\pm 0.08)\times 10^{-4}$ \\  
$\mathrm{Br}(\bar{B}_s\rightarrow D_s^+ \pi^-)$ &  $(2.98\pm 0.14)\times 10^{-3}$ \\
$\mathrm{Br}(\bar{B}^0\rightarrow D^+ K^{*-})$ & $(4.5\pm 0.7)\times 10^{-4}$ \\
$\mathrm{Br}(\bar{B}^0\rightarrow D^{*+} K^-)$ & $(2.16\pm 0.08)\times 10^{-4}$\\
$\mathrm{Br}(\bar{B}_s\rightarrow D_s^{*+} \pi^-)$ & $(1.9^{+0.5}_{-0.4})\times 10^{-3}$ \\
 \hline
\end{tabular}
\caption{Experimental values for the non-leptonic $B$ decays used in our analysis, taken from the PDG~\cite{ParticleDataGroup:2024cfk}. \label{tab:ObsBR}}
\end{table}
%
We update the experimental values for the ratios in Eq.~\eqref{eq:ratios} relative to the numbers presented in~\cite{Cai:2021mlt}, using the most recent measurements of the corresponding numerators (shown in Tab.~\ref{tab:ObsR}). These are estimated based on the branching fractions listed in Tab.~\ref{tab:ObsBR}, which are taken from the latest Particle Data Group reference~\cite{ParticleDataGroup:2024cfk} and thus include new results from Belle~\cite{Belle:2021udv,Belle:2022afp} and LHCb~\cite{LHCb:2021qbv}.
Overall this leads to slightly reduced experimental uncertainties compared to those detailed in~\cite{Cai:2021mlt} and hence increases the tension with the SM predictions.

\begin{table}[!t]
\renewcommand{\arraystretch}{1.7}  
\begin{tabular}{ c| c | c| c| c}
\hline
& Channel  & Experiment  & SM  & Pull  \\
\hline
$R_K$ &$\bar{B}^0\rightarrow D^+ K^-$ &  $0.058^{+0.004}_{-0.004}$ & $0.082^{+0.002}_{-0.001}$ & $\approx$ \qty{5.6}{\sigma} \\ 
$R_{s\pi}$ &$\bar{B}_s\rightarrow D_s^+ \pi^-$ &   $0.71\pm 0.06 $ & $1.06^{+0.04}_{-0.03}$ & $\approx$ \qty{5}{\sigma} \\
$R_{K^*}$ & $\bar{B}^0\rightarrow D^+ K^{*-}$ & $0.136\pm 0.023 $ & $0.14^{+0.01}_{-0.01}$ & $\approx$ \qty{0.16}{\sigma} \\
$R^*_{K}$ & $\bar{B}^0\rightarrow D^{*+} K^-$ & $0.064\pm 0.003$ & $0.076^{+0.002}_{-0.001}$ & $\approx$ \qty{3.6}{\sigma} \\
$R^{*}_{s\pi}$ & $\bar{B}_s\rightarrow D^{*+} \pi^-$ &  $0.52^{+0.18}_{-0.16}$ & $1.05^{+0.04}_{-0.03}$ & $\approx$ \qty{3.1}{\sigma} \\
\hline
\end{tabular}
\caption{Observables used in our low energy $B$-physics analysis, and the discrepancy between SM and experiment.\label{tab:ObsR}}
\end{table}

\subsection{Statistical Analysis}
To characterize the phenomenological imprint of new physics, we construct a $\chi^2$ function

\begin{equation}
\chi^{2}_\text{non-leptonic} = \sum_j
\Delta \vec{R}^T \cdot \vec{M} \cdot \Delta \vec{R}\,,
\label{eq:chi2_flavour}
\end{equation}

where $\Delta \vec{R}$ is a vector containing the difference between the theoretical and experimental values of the ratios appearing in Tab.~\ref{tab:ObsR}. The matrix $\vec{M}$ accounts for correlations between the different observables. It is calculated out of the experimental uncertainties and the inverse of the covariance matrix containing a sampling of the different theory computations per observable. In practice we estimate our theoretical values using Monte-Carlo sampling for each input using around 2000 evaluations per NP space point.

When comparing the current status of SM theoretical predictions for the annihilation-free non-leptonic decays described above to the latest experimental data we obtain 
\begin{equation}
\label{eq:chi2_SM_nonleptonic}
\chi^2_\text{SM, non-leptonic} = 65.5 \,.
\end{equation}
This shows a clear tension between the SM and current experimental flavour results.

\subsection{Additional flavour constraints}
\label{sec:other-flavour}
\subsubsection*{$B$ meson lifetimes}
Previous studies~\cite{Lenz:2022pgw} have highlighted the critical role that $B$-meson lifetime measurements can play in constraining four-quark operators involved in non-leptonic decay calculations.
The ratio $\tau(B^+) / \tau(B_d)$ has been measured with a precision at the sub-percent level~\cite{HFLAV:2022esi,HFLAV:2024osc}, while Standard Model predictions currently show uncertainties of around \qty{2}{\percent}. In contrast, individual $B$-meson lifetime predictions exhibit significantly larger uncertainties, typically of the order of \qty{20}{\percent}~\cite{Lenz:2022rbq}. The NP contributions from $b \to c \bar{c} (d,s)$~\cite{Jager:2017gal,Jager:2019bgk} and $b \to c \bar{u} (d,s)$~\cite{Lenz:2022pgw} operators are well-established\footnote{For convenience, we have implemented these results in Python and aim to integrate them into the public version of \flavio in the near future.}; these contributions form the second component of the total $\chi^2$ statistic we construct in Eq.~\eqref{eq:chi2_SM_nonleptonic}.
These operators also enter the prediction of the $B_s$ lifetime, which, whilst equally well-measured, is currently plagued by systematic differences related to non-perturbative input parameters as well as their dependence on $SU(3)$ flavour-breaking effects~\cite{Lenz:2022rbq}. Therefore, there is no clear avenue currently to precisely test the SM prediction against data in this observable.

Later, we will see that, in contrast to the results of~\cite{Lenz:2022pgw}, we find the lifetime ratio to not be of major importance for our overall picture.
This is due to the flavour assumptions made in \cite{Cai:2021mlt} and followed by \cite{Lenz:2022pgw}, which we do not employ -- this point will be discussed more in~\cref{sec:rge}.

\subsubsection*{\smelli likelihood}
A final group of additional flavour physics constraints on our new physics coefficients are obtained with \smelli \version{2.4.2}~\cite{Aebischer:2018iyb}, which uses \flavio \version{2.6.1}~\cite{Straub:2018kue} to calculate the BSM contributions to more than 500 observables (in the quark sector and also including Higgs physics and electroweak precision measurements amongst others) using the EFT framework. These calculations are then combined with experimental data by \smelli to provide a ``global'' $\chi^2$, which can be used to place constraints on a wide range of new physics. This forms the third part of our total~$\chi^2$.

\section{Flavour-relevant LHC constraints from the top sector}
\label{sec:lhc}
Having established the flavour constraints, we now turn to the inclusion of representative measurements of the top sector at the LHC. However, we first comment on how the flavour measurement scales need to be related to the relevant scales of the LHC collider physics to achieve a consistent combination of the data sets (see also~\cite{Englert:2014cva}).

\subsection{Connecting high and low energy physics}
\label{sec:rge}
To study the interplay of our main low energy physics observables (the non-leptonic decays) with those from the high energy top sector, we use Renormalization Group Evolution (RGE) methods to connect the SMEFT coefficients (in the Warsaw basis~\cite{Grzadkowski:2010es}) at a high scale to the coefficients of the weak effective theory at the low energy scale $\mu \sim m_b$.
Since our top constraints arise from top pair production, we choose $\mu = \qty{325}{\GeV} \approx 2 m_t$ as our high scale, turning on SMEFT coefficients at this scale. 

Firstly, we use the RGE equations for the SMEFT, which are known at leading order~\cite{Jenkins:2013zja,Jenkins:2013wua,Alonso:2013hga} to run down from our choice of high scale to the $Z$ mass. At this point, we match from the SMEFT to the low energy or weak effective theory (LEFT or WET), where the matching equations between the SMEFT Warsaw basis and the so-called ``JMS'' basis~\cite{Jenkins:2017jig} of the weak effective theory are known at both tree level~\cite{Jenkins:2017jig} and one-loop~\cite{Dekens:2019ept}. Finally, we RG-evolve again from the $Z$ mass to the $B$ scale, using the complete leading order RGE equations~\cite{Aebischer:2017gaw,Jenkins:2017dyc}. This procedure is fully implemented in the software packages \wilson \version{2.4}~\cite{Aebischer:2018bkb} and \dsixtools \version{2.1}~\cite{Fuentes-Martin:2020zaz}, which we use in our study.

\begin{figure*}[!t]
    \centering
    \includegraphics[width=\textwidth]{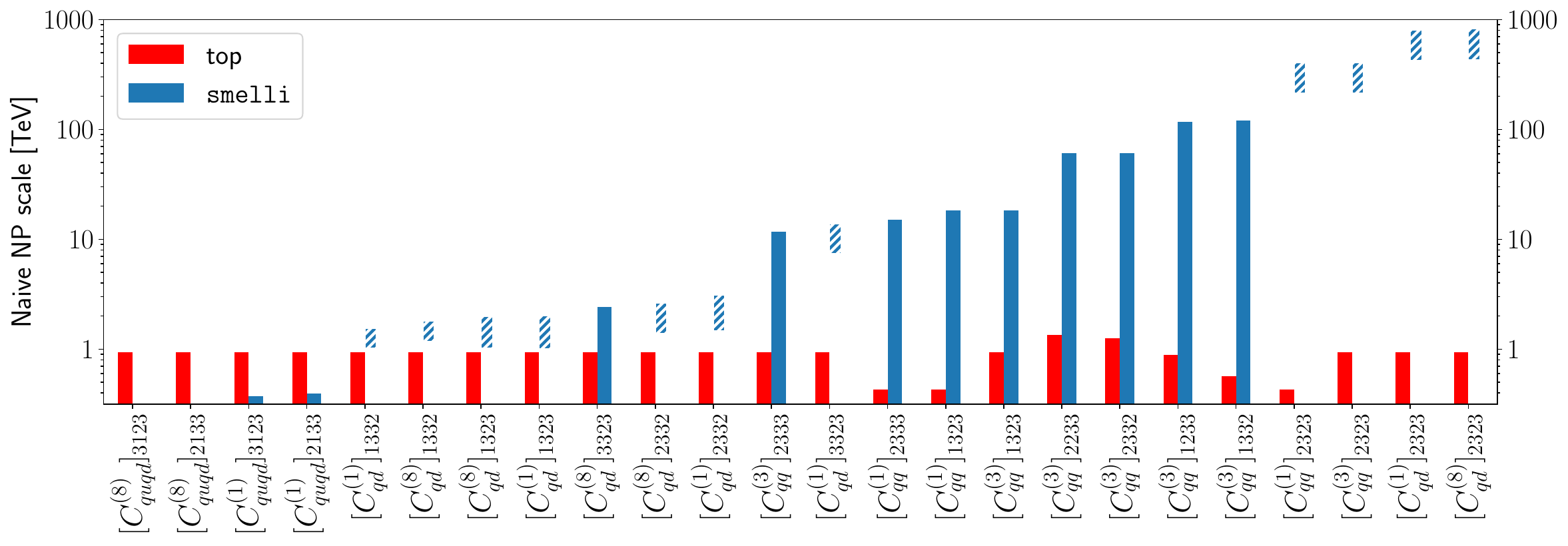}
    \caption{The lower bound on the naive NP scale corresponding to the constraints from top pair production (red) or the \smelli global analysis (blue) on operators that enter our top analysis -- where the \smelli data instead favours NP, this is instead shown as a range of NP scales (blue, hashed).
    See main text for more details.
    }
    \label{fig:top_vs_smelli}
\end{figure*}

We note that our choices differ from other works in the literature in ways relating to both the low-energy and high-energy flavour assumptions.
Firstly, for us the $b \to c \bar{u} d$ and $b \to c \bar{u} s$ sectors have no specific or fixed relation between them, in contrast to the flavour assumptions made in \cite{Cai:2021mlt} (and followed by \cite{Lenz:2022pgw} in their comparison).
Secondly, by turning on single SMEFT coefficients with specific flavour indices (see further comments in Sec.~\ref{sec:interp} below), we differ in our analysis from other studies of LHC data using SMEFT, such as \smefit~\cite{Ethier:2021bye,Giani:2023gfq} or \fitmaker~\cite{Ellis:2020unq}, where they have followed some of the flavour assumptions defined by the LHC EFT working group in Ref.~\cite{Aguilar-Saavedra:2018ksv}.

%
\begin{figure*}[!t]
    \centering
    \includegraphics[width=\textwidth]{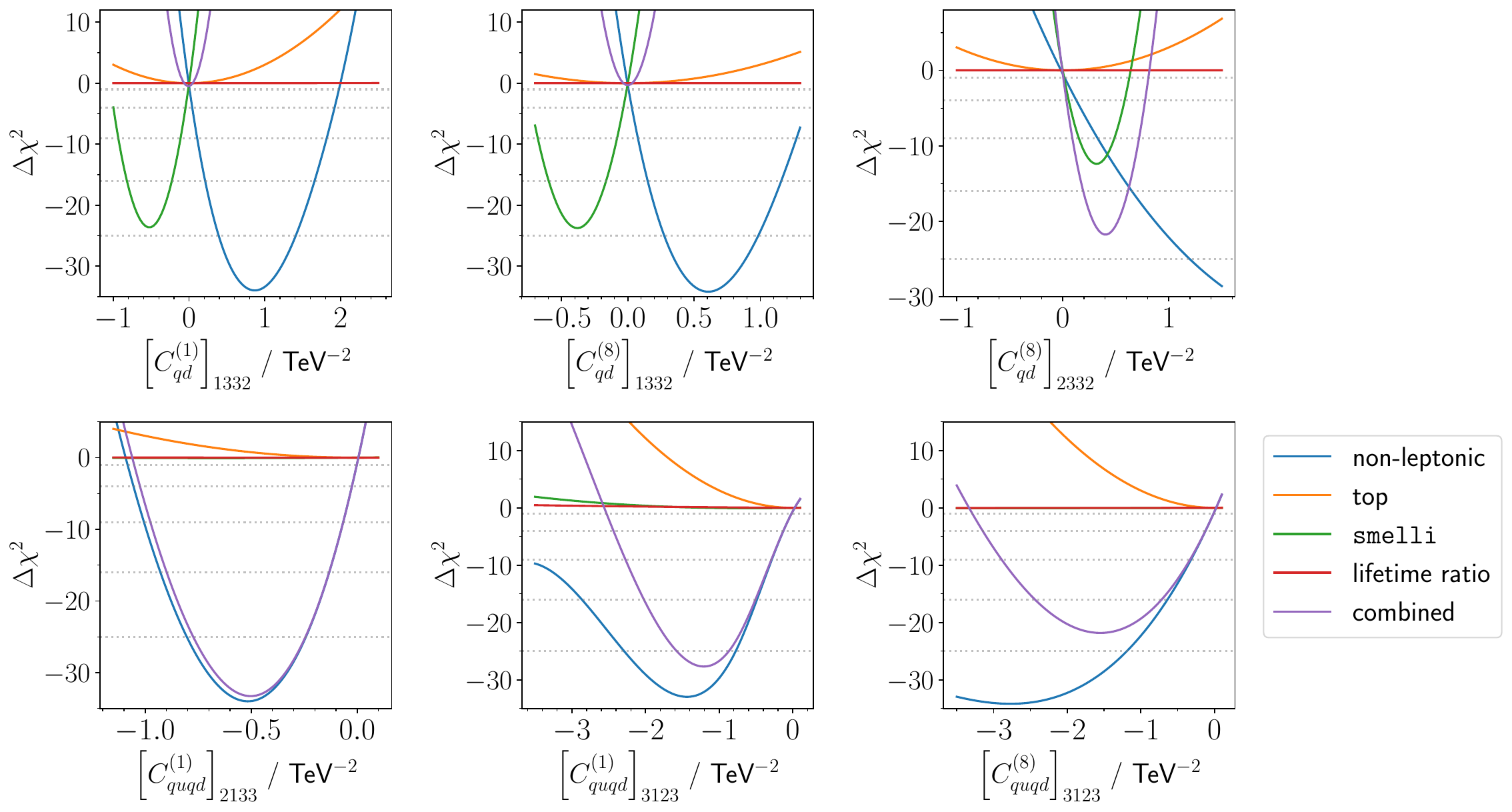}
    \caption{$\chi^2$ change for SMEFT coefficients constrained by our top analysis.}
    \label{fig:top_wc_chi2_plots}
\end{figure*}
%

\subsection{Top pair production at the LHC}
\label{sec:top}
The high-energy program at the LHC has made significant progress in honing the sensitivity to the SM production of top quarks. With a pair production cross section of around $900~\text{pb}$ (depending on the centre-of-mass energy of the proton collisions)  at the LHC, top quarks are abundantly produced at hadron colliders, which was a key motivation to consider them in the context of EFT with very early LHC data~\cite{Buckley:2015nca,Buckley:2015lku}. These efforts have intensified over the years~\cite{Brivio:2019ius,Hartland:2019bjb}, see especially the recent~\cite{Garosi:2023yxg}.

Owing to its large cross section, top pair production has recently been determined by the ATLAS collaboration at 68\% confidence level~\cite{ATLAS:2023gsl}
\begin{multline}
\sigma =829\pm1~\text{(stat)}\pm13~\text{(syst)}\\\pm 8~\text{(lumi)}\pm 2 ~\text{(beam)~pb}\,,
\end{multline}
for a luminosity of $140~\text{fb}^{-1}$ at 13~TeV. This is a 1.8\% accuracy validation of the SM expectation. In parallel, the top quark decay has been measured with an accuracy of around 10\%~\cite{ParticleDataGroup:2024cfk}.\footnote{The dominant systematic uncertainties are a function of the size of the data set, and they will likely be further reduced in the near future~\cite{Belvedere:2024wzg}.} We will take both these constraints as proxies to gauge the level of complementarity between investigations of the (comparably less precise) high energy frontier with the (comparably less energetic) precision frontier offered by flavour physics.\footnote{In fact, the constraints we will obtain quantitatively reproduce the more comprehensive analysis of Ref.~\cite{Atkinson:2021jnj} which employed the {\emph{differential}} cross section measurements by CMS~\cite{CMS:2016oae}.}

\begin{figure*}[!t]
    \centering
    \includegraphics[width=\textwidth]{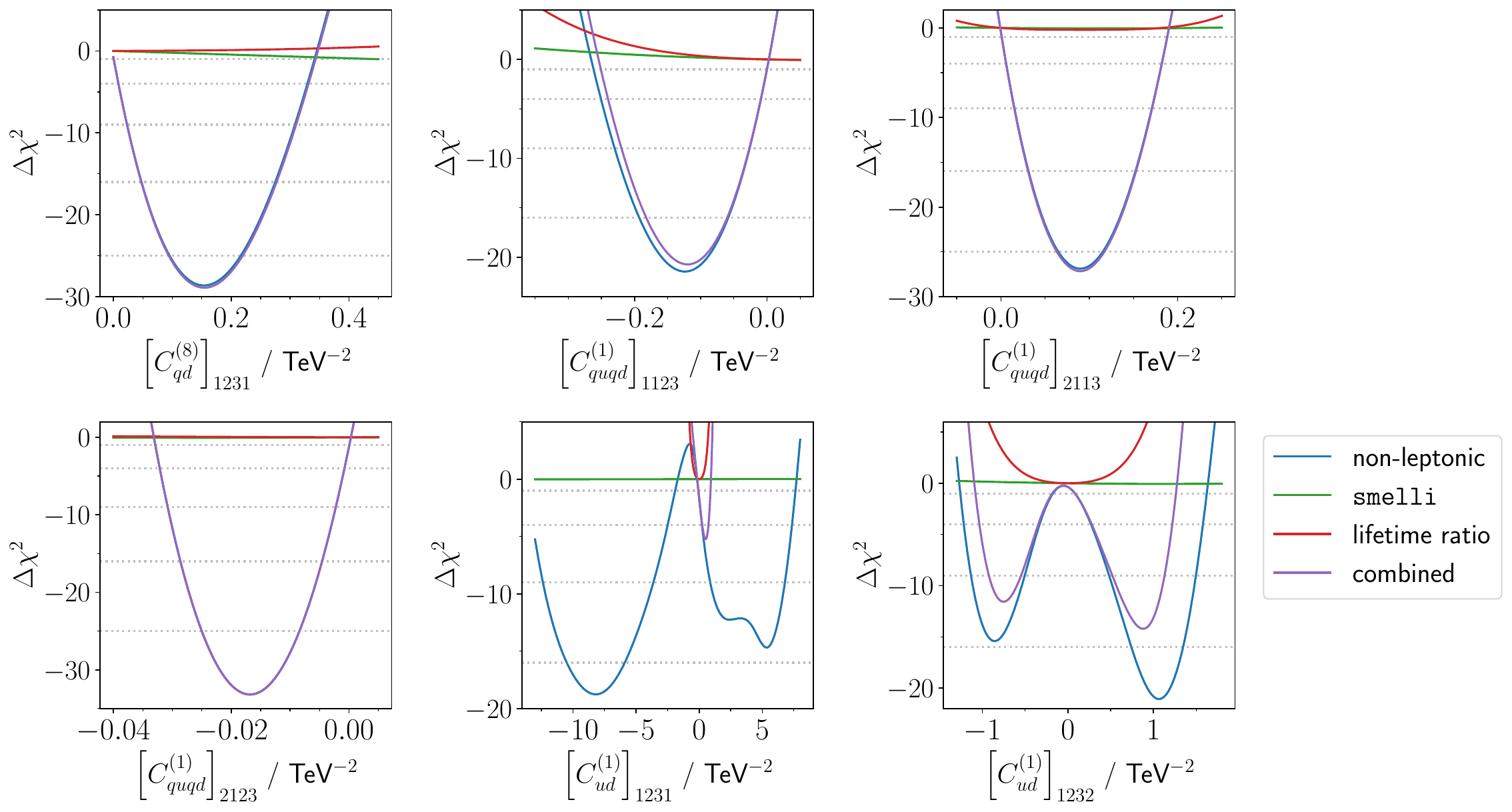}
    \caption{$\chi^2$ change for other SMEFT coefficients (not constrained by our top analysis).}
    \label{fig:other_wc_chi2_plots}
\end{figure*}

Top pair production is impacted by a range of four-fermion interactions, that are connected via the SMEFT to those we have considered in the previous section. We perform a leading order investigation of the linearized dimension-six effects of the relevant four-fermion interactions, which serves the purpose of providing a conservative estimate of the dimension-six effects for the inclusive observables in the top sector. It is well-known that squared dimension-six effects (formally these are dimension-eight) sculpt the momentum-dependent distributions and can lead to overly tight constraints (see e.g.~\cite{Buckley:2015nca,Englert:2018byk,Brivio:2019ius}). 
To obtain cross section and decay width estimates, we employ \smeftsim~\cite{Brivio:2017btx,Brivio:2020onw} and \mg~\cite{Alwall:2014hca, Hirschi:2015iia} as linear functions of the involved WCs. These results act as templates for the inclusion in a top-specific $\chi_\text{top}^2$ test statistic that we tune to reproduce the top constraints of~\cite{ATLAS:2023gsl,ParticleDataGroup:2024cfk}. Employing this $\chi_\text{top}^2$, we can analyse flavour-collider complementarity in the next section after including RGE running effects.

Already now, the experimental measurement is more precise than the precision with which we can simulate the inclusive cross section at the LHC~\cite{Czakon:2013goa}. Whilst the results of \cite{ATLAS:2023gsl} include modelling uncertainties from the Monte Carlo toolchain, these uncertainties are not actively included in the extraction of the top production cross section. Such uncertainties indeed play a role in extracting new physics parameters. As we are predominantly interested in four-fermion interactions, however, in the light of expected shape changes~\cite{Buckley:2015nca,Buckley:2015lku,Brivio:2019ius,Hartland:2019bjb,Banerjee:2020vtm,Ellis:2020unq,Ethier:2021bye,Elmer:2023wtr} these uncertainties play a subdominant role, see also~\cite{ATLAS:2022xfj}. Therefore, for the purpose of this exploratory study, we consider the cross section measurement uncertainty as provided by ATLAS and contrast it with the new physics simulation to leading order in the QCD-EFT double-series expansion and leave a more detailed differential analysis for future work
\footnote{We note that previous work \cite{Bissmann:2019qcd} has shown the importance of including experimental correlations when perform more complicated analyses using multiple experimental observables.}.

\newcommand{\tevi}{$[\text{TeV}^{-2}]$}
\begin{table*}[!t]
\renewcommand{\arraystretch}{2}  
\centering
\begin{tabular}{l|c||c|c|c|c||c}
    \hline
    Coefficient & Best-fit point~\tevi & Non-leptonic $\Delta \chi^2$ & Top $\Delta \chi^2$ & \smelli $\Delta \chi^2$ & lifetime $\Delta \chi^2$ & global $\Delta \chi^2$
    \\
    \hline
    $\left[ C_{qd}^{(1)} \right]_{1332}$   & -0.01 &   0.4 & 0   &  -0.9 & 0   &  -0.5
    \\
    $\left[ C_{qd}^{(8)} \right]_{1332}$   &  0.01 &  -1.2 & 0   &   0.9 & 0   &  -0.3
    \\
    $\left[ C_{qd}^{(8)} \right]_{2332}$   &  0.41 & -11   & 0.5 & -11.3 & 0   & -21.8
    \\
    $\left[ C_{quqd}^{(1)} \right]_{2133}$ & -0.51 & -34   & 0.8 &  -0.1 & 0   & -33.3
    \\
    $\left[ C_{quqd}^{(1)} \right]_{3123}$ & -1.21 & -32.1 & 4.4 &   0   & 0.1 & -27.6
    \\
    $\left[ C_{quqd}^{(8)} \right]_{3123}$ & -1.54 & -28.9 & 7.1 &   0   & 0   & -21.8
    \\
    $\left[ C_{qd}^{(8)} \right]_{1231}$   &  0.16 & -28.6 & - &  0.3 &  0   & -28.9
    \\
    $\left[ C_{quqd}^{(1)} \right]_{1123}$ & -0.12 & -21.4 & - &  0.2 &  0.5 & -20.7
    \\
    $\left[ C_{quqd}^{(1)} \right]_{2113}$ &  0.09 & -26.9 & - & -0.1 & -0.2 & -27.2
    \\
    $\left[ C_{quqd}^{(1)} \right]_{2123}$ & -0.02 & -33.2 & - & -0.1 &  0   & -33.3
    \\
    $\left[ C_{ud}^{(1)} \right]_{1231}$   &  0.49 &  -6.4 & - &  0   &  1.2 &  -5.2
    \\
    $\left[ C_{ud}^{(1)} \right]_{1232}$   & -0.87 & -19.1 & - & -0.1 &  5.0 & -14
    \\[0.25cm]
    \hline
\end{tabular}
\caption{The best-fit point for various SMEFT coefficients from our combined $\chi^2$ analysis, along with the breakdown of the total $\chi^2$ change in terms of our different contributions.}
\label{tab:potential_smeft_coeffs}
\end{table*}

\section{Flavour-Collider Complementarity}
\label{sec:interp}
With everything in place, we can now turn to the interplay of high-scale new physics and measurements at both low and high energy.

Our top sector analysis constrains 24 SMEFT coefficients, which shows that the top sector constraints are highly flavour-dependent. Without a definite UV model, there is no reason to prefer any particular relation between coefficients. However, a single measurement is not capable of lifting degeneracies in the BSM-extended coupling space. Hence, to highlight the qualitative interplay of the LHC and flavour measurements in this proof-of-principle analysis, we consider the WCs without marginalization. Given blind directions in the flavour and top quark EFT space~(see the discussion in Refs.~\cite{Buckley:2015lku,Englert:2017dev}, marginalization would not lead to illuminating results.\footnote{Realistic UV scenarios do typically not populate the Wilson coefficient parameter space democratically, and fully marginalized results can provide an overly pessimistic view towards realistic UV completions as a consequence.}

With this in mind, in Fig.~\ref{fig:top_vs_smelli} we compare the naive NP scale $\Lambda$ implied by the current top pair production measurements to those from the \smelli global fit, in the following way:
Using our two $\chi^2$ expressions from \smelli and top physics, we can produce a \qty{2}{\sigma} range for each of the WCs, centred around the minimum of the $\chi^2$. In cases where no NP is favoured by the data, $\chi^2_\mathrm{min}$ occurs at $C_\textrm{SMEFT}=0$ and the \qty{2}{\sigma} range has the form $-x \leq C \leq y$ ($x,y$ both positive), which we convert to a naive lower bound on the NP scale as $\Lambda_\text{NP} = \qty{1}{\TeV}/\sqrt{\max (x,y)}$. We use `max' to give a more conservative lower bound on the scale. 
In cases where NP is favoured by the data at more than \qty{2}{\sigma} significance, the \qty{2}{\sigma} range instead has the form $x \leq C \leq y$ ($x,y$ both positive or both negative), and in this case we interpret this as a naive range for the NP scale as $\qty{1}{\TeV}/\sqrt{\max(|x|,|y|)} \leq \Lambda_\text{NP} \leq \qty{1}{\TeV}/\sqrt{\min(|x|,|y|)}$.
This second case appears in our \smelli analysis given the various anomalies present in the flavour data (primarily from $b \to s \ell \ell$ transitions, see e.g.~\cite{Capdevila:2023yhq} for a summary of the current status).
From the figure we see that, while flavour measurements more strongly constrain the majority of four fermion interactions, there is a subset of interactions for which inclusive LHC top observables provide the \emph{most} competitive bounds. This sensitivity pattern arises from the distinct contributions of the operators to top pair production. The significant rates with which the tops are pair-produced at the LHC, also for the comparably smaller parton luminosities of the second generation of quarks, render the LHC constraints competitive, especially for operators that mix the first two generations. Beyond this, the constraints from the top sector are relatively flavour blind, owing to the inclusive nature of the observables we consider in this work (however, flavour tagging in the busy LHC environment is mainly limited to third-generation quarks).

How does the inclusion of top pair production impact the tensions observed in the flavour sector? By combining non-leptonic decays, top pair production, $B$-meson lifetime ratios, and global constraints from \smelli, we have identified a number of single SMEFT coefficients, which can result in a large improvement of the fit of theoretical predictions to data. Our results are collected in Tab.~\ref{tab:potential_smeft_coeffs} as well as Figs.~\ref{fig:top_wc_chi2_plots} and \ref{fig:other_wc_chi2_plots}. This clearly demonstrates the merit of looking at flavour and top data sets in tandem.
We see that for the $[Q_{qd}^{(1,8)}]_{1332}$ and $[Q_{qd}^8]_{2332}$ operators (in the language of Ref.~\cite{Grzadkowski:2010es}), the flavour data favours a non-zero value at several standard deviations, which we wish to briefly discuss. This result is driven primarily by the ongoing discrepancies in $b s \ell \ell$ observables, combined with their strong theoretical correlation with $\Delta F=2$ observables. The effect in semi-leptonic decays, starting from our four-quark operators, can be understood after examining the RG mixing -- our SMEFT WCs mix into the $[Q_{Hd}]_{23}$ coefficient. After electroweak symmetric breaking, this gives rise to a right-handed $Z$-$s$-$b$ coupling, and so upon integrating out the $Z$ we are left with an effect in the $b s \ell \ell$ sector.

Looking at the bottom row of~\cref{fig:top_wc_chi2_plots}, we see those coefficients where the pre-existing flavour constraints are weaker than our top results.
On the other hand, the top $\chi^2$ contribution is orthogonal to the non-leptonic pull. For the last two BSM coefficients considered, the current results from top measurements reduce significantly the global $\chi^2$.

Finally, we come to our results shown in~\cref{fig:other_wc_chi2_plots}, where we find more coefficients which can improve the agreement with data for the non-leptonic observables (by more than \qty{5}{\sigma} in some cases). These coefficients are, for the most part, currently unbounded by flavour physics (except for $[C_{ud}^{(1)}]_{1231,1232}$ where the lifetime ratio is strong). Top pair production is not relevant at all here, as these coefficients correspond to operators where either only a single top quark is present (and so only contribute to top decay, which is not as precise currently as pair production and so alone produces weak bounds), or the two top contribution is highly CKM suppressed, which again weakens the constraints. This motivates further study of top decay or other top observables in order to gain access to the full sensitivity potential of the LHC.

\section{Conclusions and Outlook}
\label{sec:conc}
As the LHC enters a phase of increased data-taking, the combination of high-precision low-energy measurements with statistically large datasets obtained from collisions at the highest obtainable energies will become an increasingly vital avenue to detect or exclude the presence of beyond the Standard Model physics. This, of course, is not new. In a concrete BSM extension, the inclusion of low-energy constraints is a canonical standard, and many tools that facilitate such comparisons are available. The combination of high-energy and high-precision frontiers, when approached through the agnostic lens of SMEFT, needs clarification.
This work has started~\cite{Bissmann:2020mfi,Grunwald:2023nli,Bissmann:2019gfc,Bissmann:2019qcd}. We have shown here that the LHC is becoming increasingly competitive, particularly for high statistics channels such as top pair production, where we can expect much-increased sensitivity in the near future. In parallel, we have contextualized this sensitivity with non-leptonic $B$ decay analyses (as well as more global flavour results provided through \smelli in passing) and shown how the LHC measurements constrain arbitrary flavour structures.
Obviously, within the high-dimensional space of EFT deformations of the SM, there will remain operator directions, for which either experimental setting will possess unchallenged sensitivity. However, with interactions that enable a direct combination of sensitivity across a range of energy scales, high energy physics will enter new territory of BSM exploration (see also~\cite{Bordone:2021cca,Garosi:2023yxg}).

Using the proxy LHC results linked to the top decay width and inclusive pair production cross section alongside its sensitivity to dimension-six effects, we have demonstrated how constraints interact and compare with similar analyses of the $B$ meson sector. To this end, we have shown that for a handful of SMEFT coefficients (specifically of the type $Q_{quqd}^{(1,8)}$), our LHC results are currently the \emph{most} constraining bounds available.
We have further shown that these results work against the non-leptonic observables, reducing the agreement with data at the best-fit points for new physics in our considered BSM coefficients. This highlights the importance of combining all available data when trying to identify potential directions for BSM.

Overall, our results motivate a more fine-grained analysis of the top sector, to fully monetize the increasing amount of differential information that will become available over the following years alongside similar improvements of flavour physics measurements and their interpretation. 

Particularly motivated observables here are the top quark transverse momentum and the invariant top pair mass as they directly access a potential kinematic enhancement due to four-fermion interactions. These observables will be crucial in breaking degeneracies in the SMEFT parameter space~\cite{Buckley:2015lku}. Additional processes, such as single-top~\cite{Buckley:2015lku,Degrande:2018fog} our four-top production~\cite{Zhang:2017mls,Banelli:2020iau,Aoude:2022deh} provide another arena where rare SM processes are becoming under statistical control at the LHC with high BSM potential (a qualitative overview of expected improvements at the HL-LHC have been presented recently in \cite{Belvedere:2024wzg}). Future colliders such as a FCC-hh, on the one hand, only further highlight this sensitivity potential due to a cross section increaseof all a priori sensitive SM processes when considering 100 TeV collisions. Furthermore, highly sensitive exclusive phase space regions (e.g. large invariant top pair masses) are probed in more detail. On the other hand, the precision frontier at a lepton machine such as FCC-ee can probe flavour physics aspects indirectly through an upgraded $Z$ pole programme~\cite{Allwicher:2024sso}. However, this is subject to assumptions that always underpin indirect physics analyses. Therefore, a robust improvement of the top quark programme at such a machine will crucially depend on a dedicated top quark programme at such a machine.

\bigskip
\noindent {\bf{Acknowledgments}} --- We thank Jay Howarth and Mark Owen for clarifying discussions on the ATLAS top quark physics program. We are indebted to Fang-Min Cai, Wei-Jun Deng, Xin-Qiang Li, and Ya-Dong Yang for many helpful discussions. We thank in particular Xin-Qiang Li for comparisons and very useful feedback.
Furthermore, we thank Maria-Laura Piscopo and Aleksey Rusov for their help with lifetime ratio observables. We are grateful to M\'eril Reboud for helping us with the usage of the EOS software, and to Danny van Dyk for helping with reproducing the results of \cite{Bordone:2019guc}. We thank  Thomas Mannel and Tom Tong for useful discussions.

O.A.'s work was supported by the UK Science and Technology Facilities Council (STFC) under grant ST/V506692/1.
C.E. is supported by the STFC under grant ST/X000605/1 and by the Leverhulme Trust under Research Project Grant RPG-2021-031 and Research Fellowship RF-2024-300$\backslash$9. C.E. is further supported by the Institute for Particle Physics Phenomenology Associateship Scheme and thanks the Tsung-Dao Lee Institute for hospitality during the completion of this work. 
M.K. acknowledges previous support from a Maria Zambrano fellowship, from the State Agency for Research of the Spanish Ministry of Science and Innovation through the ``Unit of Excellence Mar\'ia de Maeztu 2020-2023'' award to the Institute of Cosmos Sciences (CEX2019-000918-M), and from PID2019-105614GB-C21 and 2017-SGR-929 grants.
G.T.-X. is supported by the European Union’s Horizon 2020 research and innovation program under the Marie Sk{\l}odowska-Curie grant agreement No 945422. This research was supported by the Deutsche Forschungsgemeinschaft (DFG, German Research Foundation) under grant 396021762 - TRR 257.

\bibliography{references} 

\end{document}